# Excitation energy dependence of electron-phonon interaction in ZnO nanoparticles

Satyaprakash Sahoo[1], V Sivasubramanian, S Dhara and A K Arora

Materials Science Division, Indira Gandhi Centre for Atomic Research,
Kalpakkam 603102, India

## Abstract

Raman spectroscopic investigations are carried out on ZnO nanoparticles for various photon energies. Intensities of $E_1$-LO and $E_2$ modes exhibit large changes as the excitation energy varied from 2.41 to 3.815 eV, signifying substantially large contribution of Frohlich interaction to the Raman polarizability as compared to deformation potential close to the resonance. Relative strength of these two mechanisms is estimated for the first time in nanoparticles and compared with those in the bulk.

[1]Author to whom correspondence should be addressed;

Electronic mail: satyasahoo@igcar.gov.in

[1]Tel.: +91 44 27 480081; fax: +91 44 27 480081.

## 1. Introduction

Considerable attention has been paid to ZnO nanostructures due to their potential applications in the areas such as photodetectors, blue and ultraviolet lasers [1], optoelectronic devices, and radiation hard material for space applications [2-4]. The controlled synthesis of ZnO nanoparticles and in-depth understanding of the physical properties are the key issue for the future development of ZnO-based devices [5,6]. Zinc oxide is a wide band-gap ($E_g \sim 3.35$ eV, at room temperature) semiconductor with relatively large exciton binding energy ($\sim 60$ meV) [7]. From the point of view of electronic structure, ZnO has a strong ionic bonding. The conduction band essentially arises due to $Zn^{+2}$ $4s$ states and the upper valence band (VB) from the $O^{-2}$ $2p$ state with an admixture of $Zn^{+2}$ $3d$ states. The VB splits due to the hexagonal crystal field and spin-orbit coupling into three sub-bands A, B and C from higher to lower energy [7]. The near band-gap intrinsic absorption and emission including the defect-induced photoluminescence has been extensively studied. [8, 9, 10] On the other hand, there is significant diversity on the phonon spectra of ZnO nanoparticles.

Several resonant Raman investigation have been carried out on ZnO single crystals [11], as well as on nanoparticles [12, 13] exhibiting LO-phonon overtones up to several orders. Investigations on the confinement of phonons of different symmetries have shown [14] that the asymmetric broadening of the phonon lineshape is governed by the width of the corresponding phonon dispersion curves. Cheng *et al* [12] measured the Raman spectra of ZnO nanoparticles of sizes between 3.5 and 12 nm ( using 325 nm wavelength ) and found that $I_{2LO}/I_{1LO}$ does not vary much for this range of sizes but this ratio is much smaller than that for $\sim 30$ nm particle reported elsewhere [15]. Wang *et al* [13] studied ZnO nanowires of diameter between 20-100 nm using 325 nm wavelength. Although for such large sizes no quantum confinement effects are expected. $I_{2LO}/I_{1LO}$ was found to increase by a factor about 3 for larger diameter which is not well understood. Thus most of the studies restrict only to reporting $I_{2LO}/I_{1LO}$ ratio for different sizes without carrying any quantitative analysis of Raman intensities. Although the Raman spectra of $E_2$ and $E_1(LO)$ modes in ZnO nanoparticles are reported using 514.5 nm (2.41 eV) and 325 nm excitations [16], a systematic study of electron-phonon (e-p) interaction as a function of photon energy is not reported so far. Furthermore, neither the size nor the excitation energy dependent of Raman scattering for phonon of other symmetries in nanoparticles has been investigated. Here we report first Raman spectroscopic studies on phonons of different symmetries in ZnO nanoparticles using several excitation energies. From quantitative analysis of the Raman intensities of $E_1(LO)$ and $E_2$ phonons, relative strength of contribution of deformation potential (DP) and Frohlich interaction (FI) to the e-p interaction is estimated and compared with that in bulk.

## 2. Experimental

ZnO nanocrystalline powder is synthesized at room temperature by precipitation in nonaqueous medium using the method of Schwartz *et al* [17]. For a typical synthesis of ZnO nanocrystals, 3 g (1.8 equivalent of $OH^-$) of tetramethyl ammonium hydroxide (TMAH) is dissolved in 30 ml of absolute ethanol. Zinc acetate $[Zn(Ac)_2.2H_2O]$ of amount 2 g is dissolved in 90 ml of dimethyl sulphoxide. TMAH solution is mixed with $Zn(Ac)_2$ solution with constant stirring. ZnO nanoparticles are precipitated by the addition of about 100 ml of ethyl acetate. The precipitated ZnO is washed several times by ethyl acetate and acetone. Dry ZnO nanoparticle powder is obtained by removing the solvent at elevated temperature. In order to check the reproducibity, three batches of samples were synthesized.

ZnO nanoparticles are characterized by X-ray powder diffraction on a (STOE) diffractometer with Cu-$K_\alpha$ radiation for size determination. The Raman scattering measurement are carried out using a Spex 14018 double monochromator equipped with a $Ar^+$ laser and photomultiplier tube (Hamamatsu R943-02). Near-resonant Raman scattering measurements are performed using 325 nm line of He-Cd laser as the excitation and analysed using a double-subtractive triple-monochromator (Jovin-Yvon T64000), equipped with liquid nitrogen cooled CCD detector for recording the spectra.

**3. Results and discussion**

The X-ray diffraction pattern (Fig. 1) of as-synthesized samples agrees with the wurtzite structure (JCPDS Card no. 36-1451) and the peak broadening suggests the nanocrystalline nature of the sample. The average particle sizes calculated, using the Debye-Scherrer formula, were ~ 7.7 nm. Wurtzite structure belongs to the space group $C_{6v}^4$ with two formula units per primitive cell. Zone-center optical phonons predicted by group theory are $A_1+2E_2+E_1$. In wurtzite ZnO, the polar $A_1$ and $E_1$ modes split into longitudinal optic (LO) and transverse optic (TO) components with different frequencies due to macroscopic electric fields associated with the LO phonons [18]. Figure 2 shows Raman spectra of ZnO nanoparticle, excited using 514.5, 488 and 457.9 nm lines of a $Ar^+$ laser and 325 nm line of He-Cd laser. The spectra exhibit two prominent peaks at 438 and 582 $cm^{-1}$. Based on the reported zone-center optical phonon frequencies in ZnO [18] the peaks at 438 and 582 $cm^{-1}$ are assigned to $E_2$-high and $E_1$-LO respectively. In addition, a weak peak at 336 $cm^{-1}$ is also found for non-resonant condition; this peak can be assigned to the second-order Raman scattering arising from zone-boundary phonon $E_2(M)$ at point M in the Brillouin zone [11]. When excited under near-resonant condition using 325 nm wavelength, $E_1$-LO mode at 583 $cm^{-1}$ appears strongest as compared to other modes and exhibits overtone spectra (inset Fig. 2) up to 4-LO modes. The enhancement due to resonance can be explained based on a third order perturbation calculation for the intensity for Raman scattering, that has three matrix elements and two energy denominators [19]. When the incident photon energy is varied across an electronic transition, vanishing of energy denominators are expected to result in in- and out-resonances. Resonance Raman scattering in bulk ZnO has been well documented [11], whereas that in the nanoparticles is of considerable current interest [12, 13]. However, a quantitative analysis is still lacking.

Resonance effects in the nanoparticles can be investigated in two manners; either by changing the photon energy while keeping the electronic energy levels fixed (by keeping the particle size fixed) or by changing the particle size (there by changing the electronic energy level due to carrier confinement) and keeping the photon energy fixed. When a fixed photon energy such as 3.815 eV (325 nm) is used for ZnO nanoparticles of different sizes, the band gap increases continuously with reducing particle size and approaches the excitation energy. Thus, the reported variation of the $I_{2LO}/I_{1LO}$ ratio, which is attributed to strength of electron-phonon interaction, essentially arises due to modified resonance conditions with reducing particle size. In view of this a necessity of keeping a constant resonance condition for a meaningful comparison of the spectra of nanoparticles of different sizes has been stressed [20].

Under effective mass approximation the blue shift of the band gap due to carrier confinement in 7.7 nm ZnO nanoparticles is about 21 meV. Thus the nanoparticles band gap increases to 3.37 eV. We now examine the excitation energy dependence of the Raman intensities of $E_2$ and $E_1$ (LO) phonons. In addition to the resonance enhancement, one can also see large change in the relative intensities of $E_2$ and $E_1$-LO modes for different photon energies (Fig. 2). In order to quantify this effect, we obtain the intensities of various modes by fitting Lorentzian line shapes to the spectra. As the absolute intensities are influenced by several external factors such as incident laser power, focusing, alignment and detector response, we calculated the intensity ratio of $E_2$ and $E_1$-LO at various photon energies. Thus the contributions to Raman intensity by these external factors get cancel out. Figure 3 shows the ratio $I_{E_2}/I_{E_1-LO}$, averaged over all the samples, as a function of excitation energy. One can see that $I_{E_2}/I_{E_1-LO}$ decreases rapidly as a function of incident photon energy. We analyze the intensity ratio further to obtain insight about the e-p interaction. Raman measurements were also made using 364 nm (3.408 eV) line of $Ar^+$ laser; however, existence of large PL background prevented a quantitative analysis of Raman intensities. In view of this, the intensity ratio for this energy, taken from the reported [14] data for 6 and 4 nm size particles has been included in the Fig. 3. In addition, the intensity ratio obtained from the reported results on 4 nm ZnO nanoparticles [21] using 488 nm excitation are also shown for comparison From the standard equation of Raman intensity [19] one can see that the intensities of the phonons of different symmetries are expected to differ due to the matrix element $\langle k|H_{ep}|j\rangle$ as the e-p interaction $H_{ep}$ is different for different phonons. Here, $|j\rangle$ and $|k\rangle$ are the intermediate electronic states. As the resonance enhancement factors arising from the two energy denominators [19] is expected to be same for both phonons, the ratio of the intensities of $E_2$ to $E_1$-LO will be predominantly determined by the ratio of the corresponding matrix elements, as

$$\frac{I_{E_2}}{I_{E_1-LO}} \approx \frac{\left|\langle k|H_{ep}^{E_2}|j\rangle\right|^2}{\left|\langle k|H_{ep}^{E_1-LO}|j\rangle\right|^2} \quad . \tag{1}$$

It may be pointed out that for the polar $E_1$-LO phonon both the DP and the dipole-forbidden FI contribute to the matrix element of e-p interaction. On the other hand, for the non-polar $E_2$ mode the DP alone contributes to the matrix element. Hence, Eq. (1) can be written in terms of DP and FI as

$$\frac{H_{ep}^{E_2}}{H_{ep}^{E_1-LO}} = \frac{DP_2}{DP_1 + FI} \quad , \tag{2}$$

where $DP_2$ and $DP_1$ are the phonon-specific terms corresponding to the DP interaction for the $E_2$ and $E_1$-LO modes, respectively. Thus based on the present analysis one can say that as a first approximation, the ratio of intensities plotted in Fig. 3 represents ratio of the contributions arising from DP and FI to the e-p interaction. It may be pointed out that the intensity ratio decreases by a factor of 100 as photon energy approaches resonance. This implies that away from resonance DP dominates, whereas close to resonance the contribution of FI is largest. Furthermore, large value of $I_{E_2}/I_{E_1-LO}$ at smallest photon energy (away from resonance where FI is negligible) suggests that $DP_2$ is much larger than $DP_1$. For the sake of comparison, we also show (Fig. 3) the ratio for the bulk single crystal, estimated from the data shown in Ref. [11]. One can see that the ratios are systematically higher in 7.7 nm nanoparticles, and it further increases for smaller size nanoparticles. A decrease in e-p interaction for $E_1$–LO phonon upon reduction in size is well known for ZnO nanoparticles [12]. Hence, the present results suggest that the e-p interaction for the $E_2$ phonon decreases less rapidly with the reduction in size as compared to $E_1$–LO phonon. This may be due to the long-range nature of FI.

## 4. Conclusion

In summary, ZnO nanoparticles of about 7.7 nm diameters are synthesized using precipitation in non-aqueous medium. Raman intensities of $E_2$ and $E_1$-LO phonons are analyzed to obtain insight about electron-phonon interaction in nanocrystalline ZnO. The present results suggest that electron-phonon interaction is enhanced via Frohlich interaction as excitation energy approaches the resonance value. A comparison of the ratio of $E_2$ and $E_1$-LO intensities with that of bulk suggests that the electron-phonon interaction for $E_2$ phonon decreases less rapidly than $E_1$-LO phonon with a reduction in size.

**Acknowledgement**

We thank Ms. S. Kalavathi for the XRD patterns, Dr. C. S. Sundar for interest in the work, Dr. P. R. Vasudeva Rao for support and Dr. Baldev Raj for encouragement.

.

**Figure Caption:**

Fig. 1. Typical X-ray diffraction pattern for 7.7 nm ZnO nanoparticles.

Fig. 2. Typical Raman spectra of 7.7 nm ZnO nanoparticles using different excitation wavelength. The inset shows the overtone spectra for 325 nm excitation. Other samples exhibited similar spectra.

Fig. 3. Average intensity ratio $I_{E_2}/I_{E_1-LO}$ as a function of incident photon energy for 7.7 nm ZnO nanoparticles. The intensity ratio for 6 nm (symbol *) is taken from Ref. [14] and 4 nm (symbol Δ) are taken from Ref. [14] and [21]. For the sake of comparison, this ratio for a single crystal, estimated from the results of Ref. [11], is shown as open symbols. Dot-dash curve through the present data and dashed curve through the open circles are guide to the eye.

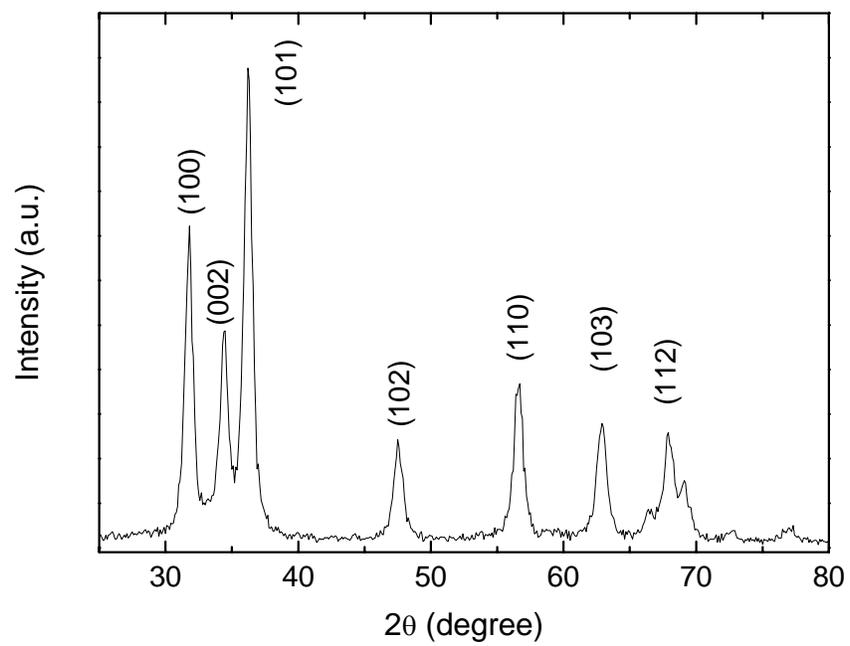

Fig. 1. Sahoo *et al*.

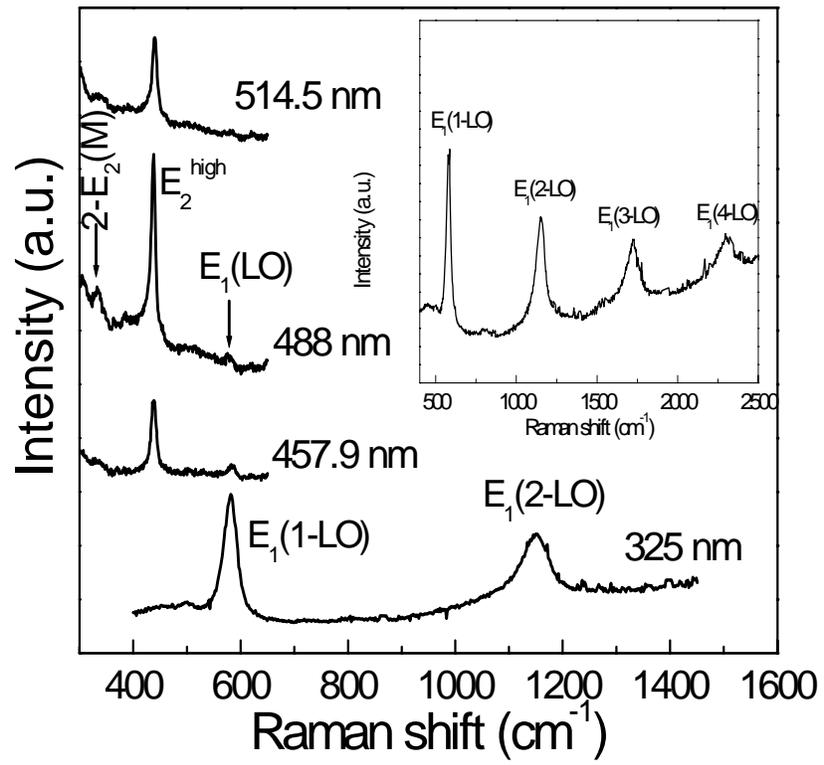

Fig. 2. Sahoo *et al*.

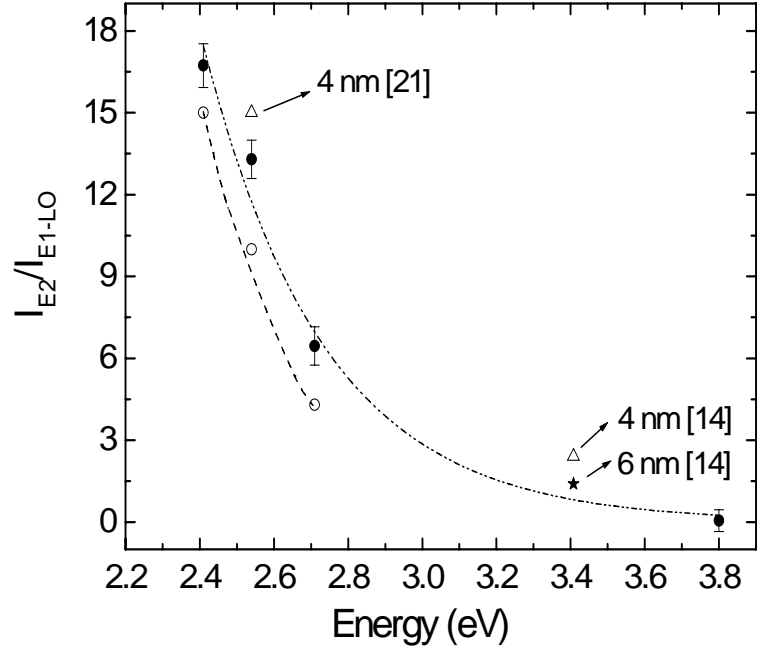

Fig. 3. Sahoo *et al*.